# NA49 evidence for the onset of deconfinement

**Volker Friese[1]**

*GSI - Helmholtz-Gesellschaft für Schwerionenforschung mbH*
*Planckstraße 1, 64291 Darmstadt, Germany*
*E-mail:* `v.friese@gsi.de`

**for the NA49 collaboration**

We review the experimental results obtained by the NA49 collaboration in the context of its beam energy scan programme. The data on particle yields and spectral distributions suggest that the deconfinement phase transition is first reached in central collisions of heavy nuclei at about 30$A$ GeV beam energy. Hadron-string transport models as well as the hadron gas model fail to describe the observed energy dependences unless additional parameters or unmeasured states are included.

---

[1] Speaker

## 1. Introduction

It is commonly believed that in central collisions of heavy nuclei at top SPS energy ($\sqrt{s_{NN}} = 17.3$ GeV) as well as at RHIC energies, a transient state of deconfined quarks and gluons is produced. In the phase diagram of strongly interacting matter, the hadronic final state of such collisions, analysed in the framework of the statistical hadron gas model, are close to the phase boundary predicted by lattice QCD calculations. The question at which collision energy the deconfinement is first reached motivated the beam energy scan programme of the NA49 experiment at the CERN-SPS, studying Pb+Pb collisions from top SPS energy down to 20$A$ GeV. The analysis of the data obtained within this programme has been finished recently; moreover, preliminary data of the STAR experiment at RHIC at comparable energies now allow to crosscheck the experimental data. In this article, we will summarise the findings of the NA49 collaboration in the context of the energy scan programme and discuss them with respect to the onset of deconfinement at lower SPS energies.

## 2. The NA49 experiment

The NA49 detector system, operated at the CERN-SPS from 1994 to 2002, is a large-acceptance hadron spectrometer, providing the measurements of identified hadron yields in a large region of phase space, covering the forward rapidity hemisphere. Because of the symmetry of the reaction, this allows to extract 4π integrated multiplicities. The tracking of charged particles is performed by four large-volume time-projection chambers operating in and after the dipole fields of two superconducting magnets. Particle identification is achieved by the measurement of the specific energy loss in the TPC gas as well as by time-of-flight measurement. Resonances are identified by the invariant-mass method; weak decays are measured through their decay topology. A detailed description of the apparatus can be found in ref. [1].

## 3. Pion yields

Pions, as the most abundantly produced particle species, measure the early-stage entropy created in the system. Figure 1 shows the pion yield in full phase space, normalized to the number of wounded nucleons, as a function of the Fermi variable $F = \left(\sqrt{s_{NN}} - 2m_N\right)^{3/4} / \sqrt{s_{NN}}^{1/4}$ characterizing the collision energy. Compared to the same observable in p+p collisions, the heavy-ion data change from suppression to enhancement at low SPS energies. Moreover, the energy dependence of the pion yield in A+A reactions changes its behaviour at SPS energies: At AGS, a linear dependence on F is observed with a slope of about 1, while from top SPS on, the slope is about 1.3. In the Statistical model of the Early Stage [2], the entropy per wounded nucleon is proportional to F, as observed in the data, with the proportionality factor depending on $N_{DOF}^{1/4}$. The observed change in slope thus is consistent with an increase by a factor of about three in the number of degrees of freedom.

The NA49 data on pion production have recently been confirmed by preliminary data of the STAR collaboration at $\sqrt{s_{NN}} = 9.2$ GeV and $\sqrt{s_{NN}} = 19.6$ GeV, which agree very well with the NA49 data points at 40$A$ and 158$A$ GeV [4].

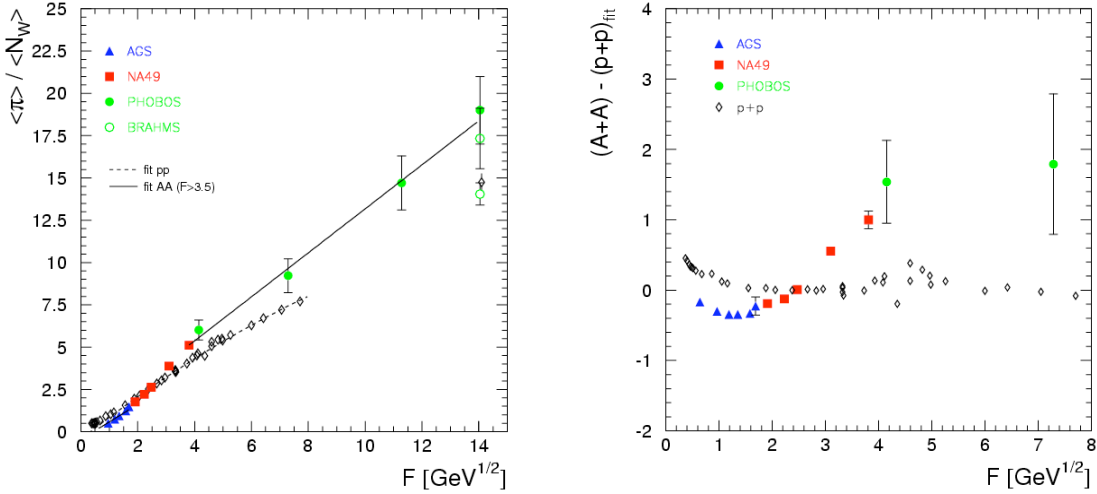

*Fig. 1. Left: pion yield in full phase space per wounded nucleon as function of the Fermi variable F in A+A and p+p collisions [3]. Right: difference of pions per wounded nucleon in A+A and p+p.*

## 4. The K/π ratio

A second striking observation is a pronounced and sharp maximum in the excitation function of the $K^+/\pi^+$ ratio at 30$A$ GeV as shown in fig. 2. Again, the low-energy STAR data are in very good agreement with the NA49 results [4]. The "horn" feature is not seen in p+p collisions, nor reproduced by hadron-string models like RQMD, UrQMD or HSD. As $K^+$ is by far the most abundant carrier of anti-strangeness at SPS energies, it provides a good measure of the total strangeness produced in the collision. The $K^+/\pi^+$ ratio thus represents the strangeness to entropy ratio. A sharp maximum in this quantity was predicted by the Statistical Model of the Early Stage as a consequence of the transition to a deconfined state [2].

A similar maximum at the same beam energy is observed also for other strange particles like Λ and Ξ⁻ [5]. This confirms that the feature is not particular to the $K^+$, but to the total strangeness content of the final state. Figure 3 (left) demonstrates this by showing the energy dependence of the total s and anti-s quark yield as derived from the measured yields by assuming isospin symmetry and correcting for unmeasured particle yields by predictions of the hadron gas model. The yields of strange and anti-strange quarks agree within errors, demonstrating the experimental consistency of the data. When normalized to the number of pions, the maximum at 30$A$ GeV is again visible (see fig. 3 (right)).

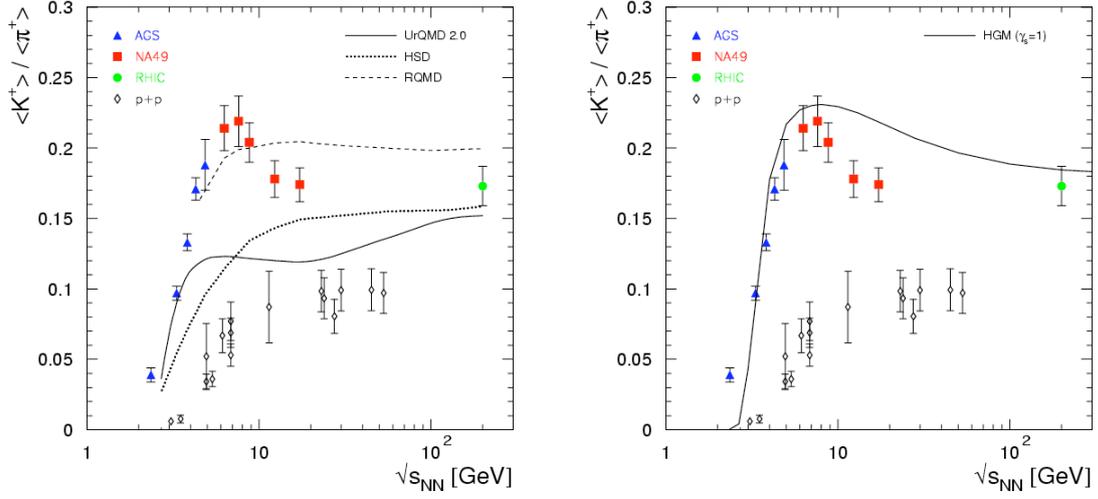

*Fig. 2. $K^+/\pi^+$ ratio in full phase space as a function of collision energy [3]. The data are compared to results of hadron-string transport models (left) and of the hadron gas model (right).*

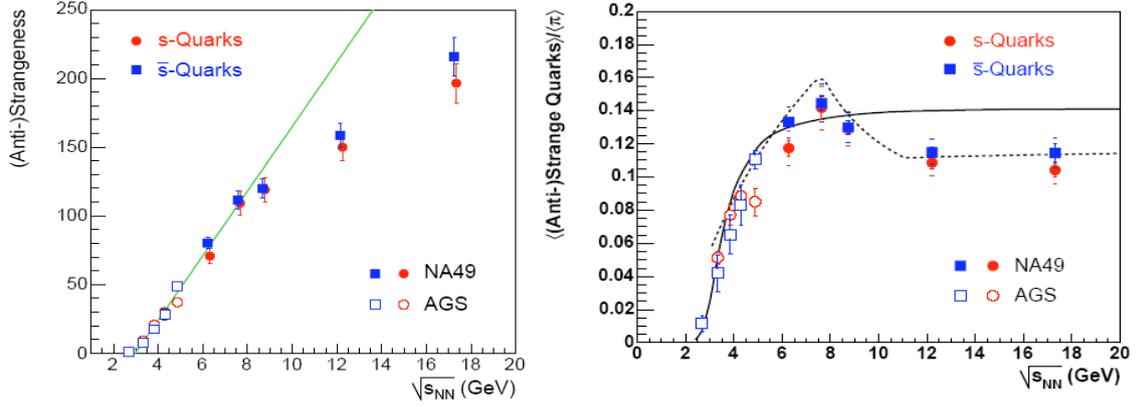

*Fig. 3. Left: Total yield of strange and anti-strange quarks as function of collision energy. Right: Strange and anti-strange quark yield normalised to the number of pions as function of beam energy*

While being overall successful in describing the final-state hadron yields, the hadron gas model in its full equilibrium version [6] does not manage to reproduce the energy dependence of the relative strangeness yields as shown in fig. 2 for the $K^+/\pi^+$ ratio. Instead, it overestimates the strange particles yields from 30$A$ GeV on. This led to the introduction of additional parameters $\gamma_s$, $\gamma_q$ accounting for the deviation from equilibrium, which indeed allows a good description of the NA49 data [7]. The fits to the hadron multiplicities result in a step-like behaviour of the parameters T and $\gamma_s$ between 20$A$ and 30$A$ GeV, indicating a sudden change in the underlying production mechanism in this energy region.

A different approach to cure the deficits of the hadron gas model in reproducing the measured strangeness yields is the inclusion of high-mass resonances in the model spectrum [8, 9]. Since these resonances feed predominantly into pions, the K/π ratio is reduced at higher energies where the temperature parameter saturates. An even better description is reached when a full Hagedorn spectrum is tentatively included. This extension of the hadron gas model, however, relies on unmeasured hadronic states and educated-guess assumptions on the branching ratios.

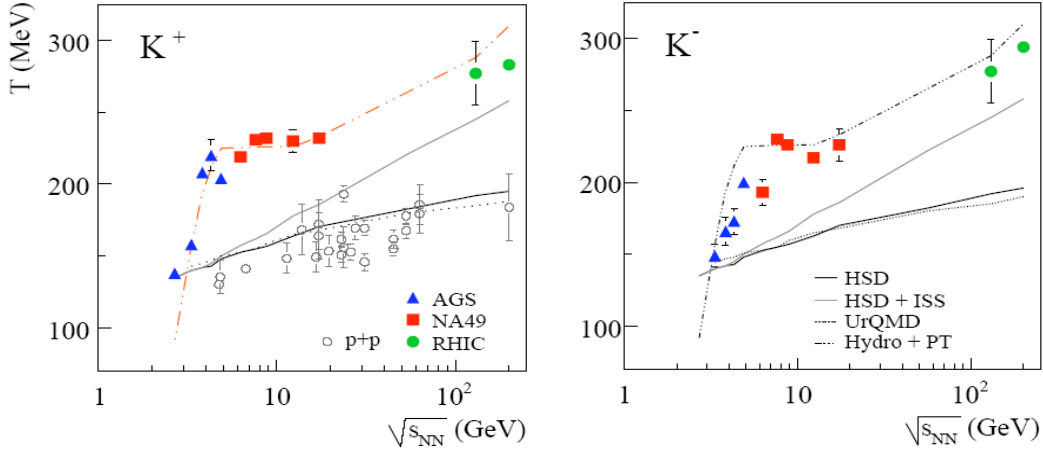

*Fig. 4. Energy dependence of the inverse slope parameter for (left) $K^+$ and (right) $K^-$ [3]. The data are compared to results of UrQMD, HSD and a hydro-.dynamical model including a first-order phase transition [10].*

## 5. Transverse mass spectra

More evidence for the onset of the transition to a deconfined state is obtained from the kinetic distribution of the final-state hadrons. The transverse-mass spectra of kaons are in good approximation exponential. Figure 4 shows the inverse slope parameters for $K^+$ and $K^-$ as a function of collision energy, which are found to be constant over the entire SPS energy range, while rising strongly at the lower SPS energies and gradually from top SPS to RHIC. Again, the NA49 data are in excellent agreement with STAR results at low energies [3, 11]. As in the case of kaon yields, this behaviour is not seen in p+p reactions nor reproduced by transport models.

The slope parameters measure both the local temperature and the pressure of the expanding system. Thus, the step-like behaviour observed for the kaon slopes is consistent with constant temperature and pressure in the mixed phase of a first-order phase transition. Indeed, a hydrodynamic model including such a phase transition provides a satisfying description of the experimental data [10].

As fig. 5 demonstrates, the feature is not particular to kaons, but also present in the transverse mass spectra of pions and protons. Here, we choose the average transverse mass to characterise the spectra because of their non-exponential form.

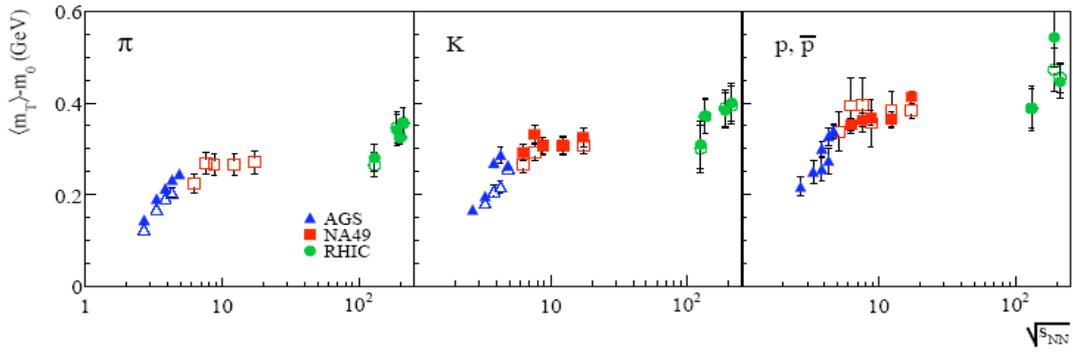

*Fig. 5. Average transverse mass of pions, kaons and protons as function of collision energy*

## 6. Rapidity distributions

The large acceptance of the NA49 detector offers the unique possibility to study the rapidity distributions of the final state hadrons. We find a scaling of the widths of the rapidity distributions with beam rapidity as shown in fig. 6. At each energy, a mass ordering of the widths is observed, with the pions as the lightest particles having the largest rapidity width.

In the Landau hydro-dynamical model, the rapidity width can be related to the sound velocity $c_s$. When applying this prescription to the NA49 data, a minimum of $c_s$ is found at low SPS energies (fig. 6 (right)), indicating a softest point of the equation of state as expected in the case of a first-order phase transition [12]. This observation is thus consistent with the conclusions derived from pion and kaon yields and the transverse mass spectra.

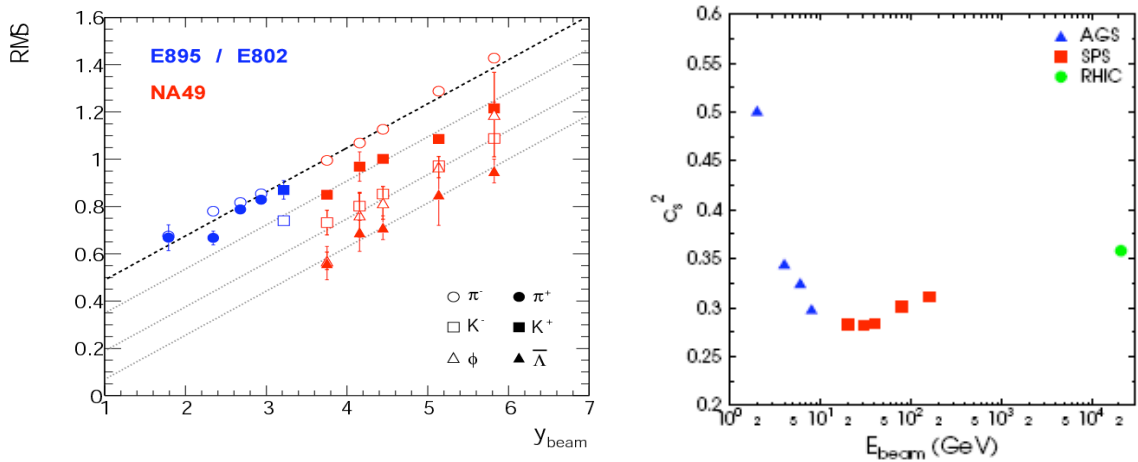

*Fig. 6. Left: Widths of the rapidity distributions of pions, kaons, Lambda and phi as function of beam rapidity in the lab system. Right: Energy dependence of the square of the sound velocity, derived from the pion rapidity width [12].*

## 7. Bose-Einstein correlations

Seemingly in contradiction to the observations described above, the radii derived from the Bose-Einstein correlation functions of charged pions [13] show a remarkably small if any change with collision energy from AGS to RHIC. In the case of a first-order phase transition, a long duration of pion emission is expected, which would manifest in $R_{out} \gg R_{side}$. No indications of this are seen in the data. It should be noted, though, that the interpretation of the HBT radii remains a puzzle since even at RHIC, they are not reproduced by hydro-dynamics. A possible way out of this has recently be proposed [14].

In conjunction with the transverse mass spectra, the Bose-Einstein correlations allow to extract the average phase-space density of negatively charged pions [15]. This quantity exhibits a plateau at SPS energies (see fig. 7) similar to that of the average transverse mass, which may be associated with new physics setting in at low SPS energies.

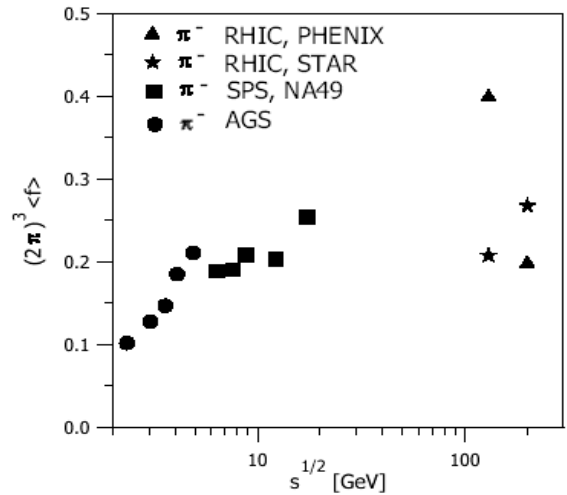

Fig. 7. Energy dependence of the average phase space density of $\pi^-$, derived from HBT data and transverse mass spectra [15]

## 8. The $\phi$ meson

Consisting of a pair of strange and anti-strange quarks, the $\phi$ meson plays an important role in the understanding of strangeness production, albeit not contributing significantly to the total number of produced s and anti-s quarks. Figure 8 shows the $\phi$ yield in full phase space, normalised to the average pion multiplicity, as a function of collision energy. Evidently, neither UrQMD nor the full equilibrium hadron gas model provide a good description of the measured data. However, in case of UrQMD, this failure is mostly due to the overestimation of the pion yield; the $\phi$ yield itself is reasonably well reproduced with the exception of the data point at top SPS energy. Recalling that UrQMD produces about 70 % of all $\phi$ mesons by coalescence of kaons, we compare the rapidity widths of ϕ with that of $K^+$ and $K^-$ (fig. 9). Only at the lowest SPS energy (20$A$ GeV), the observed ϕ width agrees with that expected from kaon coalescence. We conclude that there is no hadronic description of ϕ meson production from low SPS energies on.

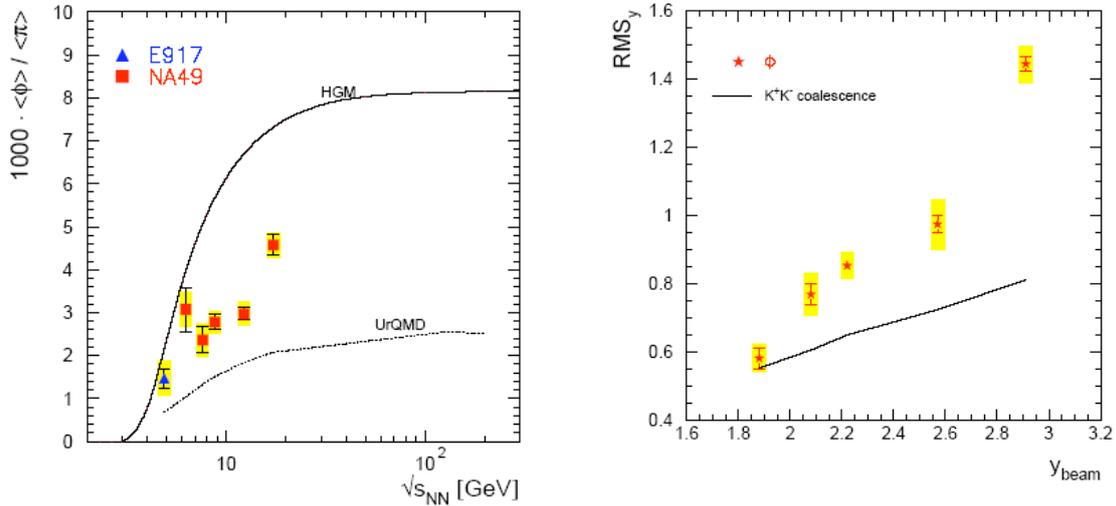

*Fig. 8. Left: energy dependence of the ϕ/π ratio in full phase space [16] compared to the prediction of UrQMD and the full equilibrium hadron gas model. Right: Energy dependence of the ϕ rapidity width compared to the expectation for the kaon coalescence picture.*

## 9. Summary

The observations obtained by the NA49 collaboration in the context of its beam energy scan programme strongly suggest that deconfinement first sets in at low SPS energies (20$A$ – 30$A$ GeV). The observed features – the change of the energy dependence of the number of pions per wounded nucleons, the sharp maximum in relative strangeness production, the plateau in the average transverse mass and in the average phase space density of pions, the minimum in the sound velocity – are particular for heavy-ion collisions and not present in elementary reactions like p+p. They are reproduced neither by hadron-string transport models nor by the hadron gas model, unless additional parameters or unmeasured states and branching ratios are included in the latter. On the other hand, the experimental data can be well understood within models that assume a first-order phase transition at about 30$A$ GeV.